\documentclass{bmcart}

\usepackage[utf8]{inputenc} 

\def\includegraphics{}

\startlocaldefs
\usepackage{amsmath,amssymb,mathrsfs,amsfonts,dsfont}
\usepackage{graphics,graphicx}
\usepackage{fancyhdr}
\usepackage{float}
\usepackage{multirow}
\usepackage{enumerate}
\usepackage{url} 
\usepackage{hyperref}
\usepackage{breakurl}
\usepackage{bm}
\usepackage{xcolor}
\endlocaldefs

\begin{document}

\begin{frontmatter}

\begin{fmbox}
\dochead{Research}


\title{ProgPerm: Progressive permutation for a dynamic representation of the robustness of microbiome discoveries}


\author[
   addressref={aff1},                   
         corref={aff1},                       
   email={liangliangzhang.stat@gmail.com}   
]{\inits{LZ}\fnm{Liangliang} \snm{Zhang}}
\author[
   addressref={aff2},
   email={shiyushu2006@gmail.com}
]{\inits{YS}\fnm{Yushu} \snm{Shi}}
\author[
addressref={aff1},
email={kimdo@mdanderson.org}
]{\inits{KAD}\fnm{Kim-Anh} \snm{Do}}
\author[
addressref={aff1},
    noteref={n1},                        
email={CBPeterson@mdanderson.org}
]{\inits{CBP}\fnm{Christine B.} \snm{Peterson}}
\author[
addressref={aff3},
         noteref={n1},                        
email={RRJenq@mdanderson.org}
]{\inits{RRJ}\fnm{Robert R.} \snm{Jenq}}

\address[id=aff1]{
  \orgname{Department of Biostatistics, University of Texas MD Anderson Cancer Center}, 
  \city{Houston},                          
  \state{Texas} ,                          
  \cny{USA}                                    
}

\address[id=aff2]{%
	\orgname{Department of Statistics, University of Missouri},
	\city{Columbia},
	\state{Missouri},
	\cny{USA}
}

\address[id=aff3]{%
  \orgname{Department of Genomic Medicine, University of Texas MD Anderson Cancer Center},
  \city{Houston},
  \state{Texas},
  \cny{USA}
}


\begin{artnotes}
\note[id=n1]{Equal contributor} 
\end{artnotes}

\end{fmbox}


\begin{abstractbox}

\begin{abstract} 
\parttitle{Background} 
Identification of features is a critical task in microbiome studies that is complicated by the fact that microbial data are high dimensional and heterogeneous. Masked by the complexity of the data, the problem of separating signals from noise becomes challenging and troublesome. For instance, when performing differential abundance tests, multiple testing adjustments tend to be overconservative, as the probability of a type I error (false positive) increases dramatically with the large numbers of hypotheses. 
Moreover, the grouping effect of interest can be obscured by heterogeneity. These factors can incorrectly lead to the conclusion that there are no differences in the microbiome compositions.

\parttitle{Results} 
We translate and represent the problem of identifying differential features as a dynamic layout of separating the signal from its random background. 
We propose progressive permutation as a method to achieve this process and show converging patterns.   
More specifically, we progressively permute the grouping factor labels of the microbiome samples and perform multiple differential abundance tests in each scenario. We then compare the signal strength of the top features from the original data with their performance in permutations, and observe an apparent decreasing trend if these top features are true positives identified from the data. We have developed this into a user-friendly RShiny tool and R package, which consist of functions that can convey the overall association between the microbiome and the grouping factor, rank the robustness of the discovered microbes, and list the discoveries, their effect sizes, and individual abundances. 

\parttitle{Conclusions} 
Simulations and applications on real data show that the proposed method creates a U-curve when plotting the number of significant features versus the proportion of mixing. The shape of the U-Curve can convey the strength of the overall association between the microbiome and the grouping factor. We also define a fragility index to measure the robustness of the discoveries. Finally, we recommend the best features by comparing p-values in the observed data with p-values in the fully permuted data.

\end{abstract}




\end{abstractbox}
%

\end{frontmatter}



\section*{Background}

With the advent of next-generation sequencing technologies to quantify the composition of human microbiome, there have been drastic increases in the number of microbiome studies and vast improvements in microbiome analysis~\cite{knight2018best}. In recent decades, a tremendous amount of evidence has strongly suggested that the human microbiota is becoming a crucial key to understanding human health and physiology~\cite{jie2017gut,vogt2017gut,cani2018gut,wei2018xiexin,gopalakrishnan2018influence,ong2018gut,riquelme2019tumor}. In practice, identification of microbial biomarkers often requires singling out specific taxa that are differentially abundant between two groups of interest (e.g. treatment vs. control). Differential abundance analysis~\cite{paulson2013differential} in this setting, however, is challenging. {On the one hand, microbiome data are high dimensional with complex structures.  A single sample can produce as many as tens of thousands of distinct sequencing reads. These reads are clustered into operational taxonomic units (OTUs) and mapped to the microbial species according to a reference library. At the same time, the OTUs (which can be considered as the lowest level taxa) are routinely aggregated to higher taxonomic levels (phyla, order, class, family, genus, or species). On the other hand, microbiome data are heterogeneous across subjects that belong to different populations, because microbiome samples interact with different body environment that might be depicted by multiple clinical outcomes. It is highly likely that not all of these host phenotypes are collected and included in the study, but with all the available clinical factors in the current data, we would like to explore and investigate a subset that are most associated with differences in microbiome compositions. Then we would like to identify the corresponding microbiome features that are significantly and robustly associated with these clinical outcomes. }

Researchers have adapted classical differential analysis tools developed for RNA sequencing data, such as edgeR~\cite{robinson2010edger} and DESeq~\cite{love2014moderated}, to microbiome data, as both data types are essentially read count data. Others have proposed methods that account for the compositional nature of microbiome data, including ANCOM~\cite{mandal2015analysis} and ALDEx2~\cite{fernandes2014unifying}. 
Segata N, et al.~\cite{segata2011metagenomic} developed LEfSe (Linear discriminant analysis Effect Size) to identify differential taxonomic features between groups by using standard tests for statistical significance. When doing multiple tests, the probability of a Type I error (false positive) increases dramatically as high throughput sequencing data is tested~\cite{goeman2014multiple}. Adjustment methods such as the Benjamini-Hochberg procedure will become over-conservative and incorrectly lead to conclusions that there are no differences in the microbiome, because the threshholds of rejecting the null hypothesis for each microbe becomes extremely small as the number of tests increases~\cite{jiang2017discrete}. Although these differential testing methods are able to identify the significance of individual microbiomarkers when associating with a single clinical outcome, they do not answer a more general question as to which grouping factors better identify more differences in microbiome communities and deserve further analysis when multiple clinical outcomes are presented in the observed data. Researchers usually use dimension reduction plots (e.g. PCoA or NMDS) at the beginning to explore the overall associations between clinical outcomes and microbiome compositions before any further investigations. {But the expected clustering effect may or may not be observed depending on the degree of heterogeneity across samples and populations, which could lead to the false conclusion that the microbiome is not associated with a clinical factor. Therefore, a systematic tool is needed to explore both the overall and the individual associations, and to provide measures on the robustness of the discoveries and the reliability of the results. }

We propose a novel method named progressive permutation. 
The method progressively permutes the grouping factor labels of microbiome samples and performs differential testing (such as a Wilcoxon rank-sum test or a Kruskal-Wallis test) on the permuted data in each scenario. We then compare the signal strength ($-\log_{10} p$-values) of top hits from the observed data with their testing performance in permuted data sets. We can observe an apparent decreasing trend of the signal strength from the no permutation scenario to the full permutation scenario, if these top hits are true positives identified from the data. As the fragility index is a measure of the robustness of the results of a clinical trial~\cite{walsh2014statistical, feinstein1990unit}, we propose a similar concept in our progressive permutation to measure the minimum number of permutation steps that would change the variable's significance to nonsigificance. We also extend these concepts to a continuous outcome using correlation tests (such as Kendall's tau or Spearman Rank Correlation tests). We have developed this method into a user-friendly and efficient RShiny tool with visualizations, so that the method becomes easy to apply, the results are easy to understand and the process of analyzing is well organized. Hawinkel S, et al.~\cite{smirnova2019perfect} proposed a permutation filtering method to measure the taxa importance by the filtering loss of exclusion of the taxa. The method randomly permutes the labels of taxa and evaluates the proportion of total variance loss. Our method permutes the sample labels to regroup them and evaluate the robustness of group differences. We validate our method with simulations and applications in real data. {We conclude that the proposed method can not only compare the overall association between the microbiome and multiple grouping factors (that might be obscured by heterogeneity), but also single out the robust individual hits. It achieves the former by 
measuring the changing trend of the number of significant hits across permutation scenarios and ranking the fragility index of the discovered microbes. It achieves the latter by comparing the p-values of the observed data (signals) with p-values of the fully permuted data (noise). To finalize the results, the RShiny tool lists the discoveries, their effect sizes and individual abundances.} 

The paper is organized as follows. In Section 2, we include a detailed description of the proposed method. In Section 3, we run simulations, and use the U-Curve and fragility index to measure overall associations with grouping factors and the robustness of microbiome discoveries. In Section 4, we apply the method to real data to test overall associations and identify robust hits. In section 5, we show the analytical properties of the proposed method in a simple setup. We conclude with a discussion in Section 6.


\section*{Methods}

Suppose that we collect $N$ samples and obtain $p$ microbiome taxa. We denote the microbial features as $\bm X=(\bm x_1,\cdots, \bm x_p)$, where each $\bm x_i$ is an $N$-dimensional vector. We aim to identify which variables are differential by the grouping factor of interest with two groups $\bm g=(\bm g^1,\bm g^2)$. We denote the grouping labels in group 1 as $g^1_i=1, i= \{1, \cdots, n_1\}$ and group 2 as $g^2_i=2, i= \{1, \cdots, n_2\}$, where $n_1+n_2=N$. The hypothesis test performed on each variable is denoted as $H_j, j=\{1,\cdots, p\}$. The corresponding p-value is denoted as $p_j, j=\{1, \cdots, p\}$. 

We use $k=\{0, 1, \cdots, K\}$ to describe progressive permutation scenarios. $k=0$ describes the observed data without any permutation. $K=\mathrm{min}(n_1, n_2)$ is the maximal permutation scenario.  
The permutation scenario $k$ is constructed as follows. Each time, we start from the original grouping labels $\bm g=(\bm g^1,\bm g^2)$. We randomly draw $k$ samples from group $1$ (sample labels $\{i_1^1,\cdots,i_k^1\}\subseteq \{1, \cdots, n_1\}$) and $k$ samples from group $2$ (sample labels $\{i_1^2,\cdots, i_k^2\}\subseteq \{1, \cdots, n_2\}$), and then exchange their grouping labels, meaning that $g^1_{i}=2, i=\{i_1^1,\cdots,i_k^1\}$ and $g^2_{i}=1, i=\{i_1^2,\cdots, i_k^2\}$. In the $k$-th permutation scenario, we have $\binom{n_1}{k}\binom{n_2}{k}$ choices. The number of choices $\binom{n_1}{k}\binom{n_2}{k}$ approaches its maximum, when $k$ equals the closest integer greater than $\frac{n_1n_2-1}{n_1+n_2+2}$. We call it as the full permutation scenario with $K_f=\lceil\frac{n_1n_2-1}{n_1+n_2+2}\rceil$. If $n_1=n_2=n$, then $K_f=\lceil\frac{n-1}{2}\rceil$. Adding up the choices of all the scenarios, we get the following equation
\begin{equation}\label{eq:combn}
\sum_{k=0}^{K} \binom{n_1}{k}\binom{n_2}{k} =\binom{N}{K}.
\end{equation}
The above equation can be derived from Vandermonde's convolution identity for binomial coefficients. The details are shown in Section S1 of Supplementary material. The left side lists all the progressive permutation scenarios which are disjoint meaning that grouping labels are distinct between scenarios. The right side lists all possible combinations when you group $N$ samples into two subgroups with $n_1$ and $n_2$ samples respectively. With the increase of $k$, the two groups are mixing more with each other.  In other words, among all the grouping assignments at random, the permuted assignments more similar to the original data (the observed grouping factor) would differentiate the two groups more than the less similar ones, if the microbiome variables were strongly associated with the observed grouping factor.

Given the properties, next we introduce how to summarize the progressive permutation results to access the overall associations. For each draw in every scenario $k$, we perform $p$ independent tests to differentiate each microbiome features between the two groups and calculate all the $p$-values.  We can obtain the number of significant taxa as $\mathrm{nsig}(k)=\sum_{j=1}^p I_{p_j(k)\leq \alpha}$, where $\alpha$ is the prespecified significance level (default value is 0.05). We expect to see the lowest $\mathrm{nsig}(k)$ in the fully mixing scenario $K_f$. 
The number of significant features $\mathrm{nsig}(k)$ decreases with the proportion of mixing $k/K$, when $k\leq K_f$. $\mathrm{nsig}(k)$ increases with the proportion of mixing $k/K$, when $k\geq K_f$. 
Therefore, the number of significant features $\mathrm{nsig}(k)$ is a U-Curve of the proportion of mixing $k/K$, when $0\leq k\leq K$.
As the shape of U-Curve measures the signal strength of the microbiome compositions that are differential with the grouping factor, we potentially can use the U-Curve as a global measure to depict the overall association between microbiome compositions and multiple clinical outcomes.

\begin{figure*}[!ht]
	\centering
	\includegraphics[width=0.9\linewidth]{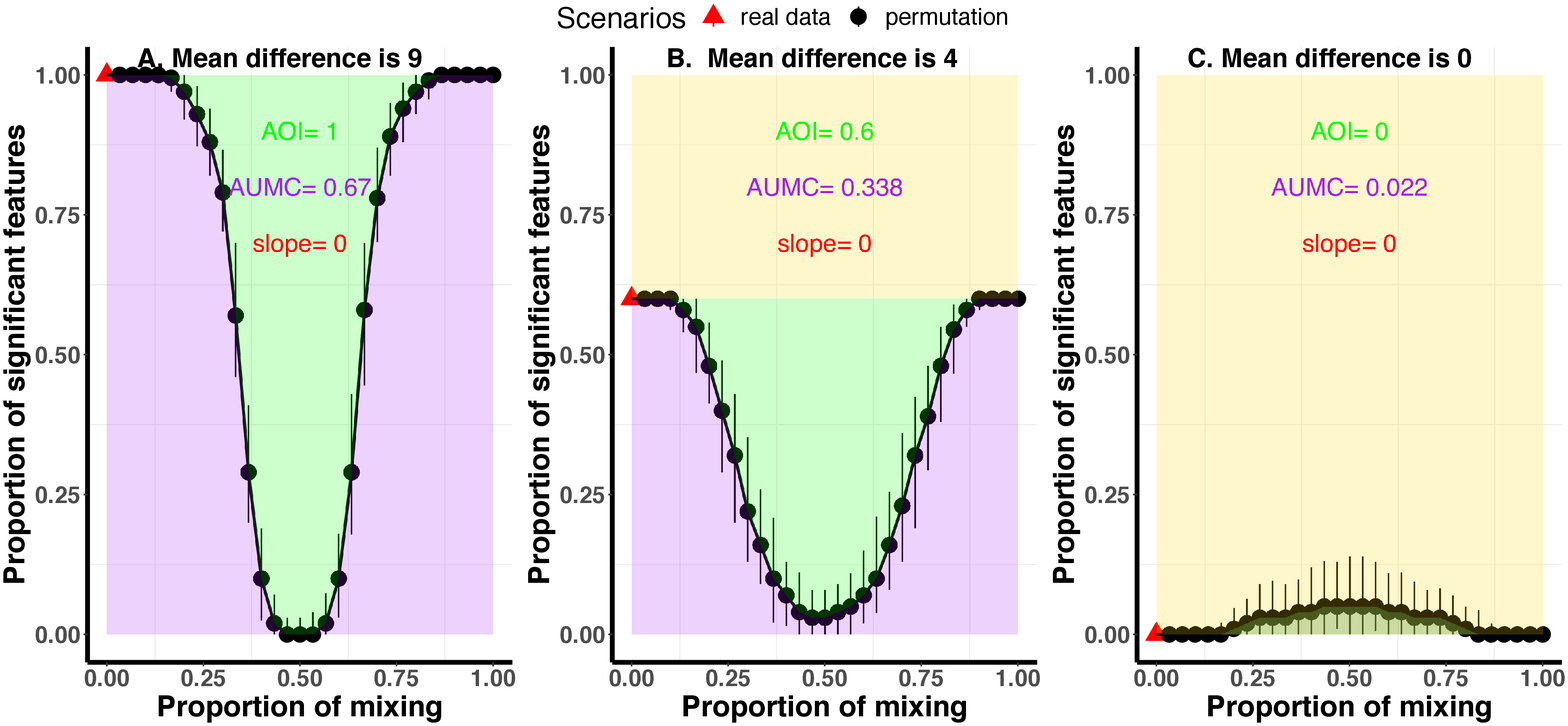}
	\caption{An illustration plot of the U-Curve of proportion of significant features versus proportion of mixing. x-axis describes the proportion of mixing the two groups of data. y-axis describes the proportion of significant features. The red triangle describes the observed data. The black dots describe the permuted data. The vertical bars describe the 95\% quantile confidence intervals.
	}.
	\label{fig:examplecurve}
\end{figure*}

{
Each permutation scenario consists of multiple combination choices, implemented as follows.
For each permutation scenario $k$ ($k\geq 1$), we start from a random seed and perform a subset of $\nu=N\left(\log\binom{n_1}{k}+\log \binom{n_2}{k}\right)$ draws. Therefore,  for each variable $j$, we obtain $\nu$ samples of  p-values $p_{j}(k)$.
We summarize the distribution of these samples by their medians $p^m_j(k)$ and 2.5\%-97.5\% quantile intervals. To visualize these p-values in an organized manner, we rank the significance of all the variables in the observed data, and then plot their median $-\log_{10} p$-values with the same order across permutation scenarios. In general, the paralleled traces of median $-\log_{10} p$-values of more significant variables will be higher than those for less significant ones. With the increase in mixing, the significant p-values gradually become nonsignificant, indicating that the signal is weaker and the noise is stronger. As there would be almost no signal if the data were fully mixed, more p-values are close to 1 at the full permutation scenario $k=K_f$. Therefore, we could observe a decreasing trend in the number of significant features from no permutation scenario ($k=0$) to the full permutation scenario $k=K_f$. If the two groups have balanced sample sizes (i.e. $n_1=n_2$), we will visualize a symmetric U-shape curve if we plot the number of significant features with the proportion of mixing $k/K$. We scale the number of significant features by total number of features considered. As illustrated in Fig~\ref{fig:examplecurve}, we define the area of interest (AOI) as the rectangular region covering the curve (green plus purple), which actually measures the proportion of significant features out of all features. To describe the shape of the U-Curve, we define area under the mixing curve (AUMC) and the decreasing slope of the initial point depicting the observed data. The slope of the initial point is calculated as the slope of the line connecting the first two points. Smaller AOI and AUMC indicates the lower association between clinical outcomes and microbiome compositions. For two clinical outcomes whose AOIs are equal, if one outcome provides smaller slope and bigger AUMC, we will conclude that the overall association between this outcome and the microbiome features were higher.
}

{
The fragility index was originally defined as a measure of the robustness of the results of a clinical trial~\cite{walsh2014statistical, feinstein1990unit}. We introduce a similar concept to measure how fast the signals break down as the mixing increases. We introduce and define the fragility index of $j$th variable of each draw at permutation scenario $k$  as $\mathrm{FI}_j=\min_k \left(p^m_j(k)>\alpha\right)$, where $p^m_j(k)$ is the median p-value obtained above in each scenario $k$. In other words, the fragility index of a variable is the minimum number of permutation steps that would change the variable's significance into nonsignificance. {So the fragility index is less than full permutation scenario $K_f$, where all p-values are not significant. Therefore, we can obtain the scaled fragility index as $\mathrm{sFI}_j=\mathrm{FI}_j/K_f$.} The larger the fragility index is, the more stable the identified taxa are. Therefore, within the same data set, we can rank the importance of the taxa by their fragility indices. For two clinical outcomes, if one outcome is more associated with microbiome features, this outcome will provide higher average fragility indices. 
	
If we roll back the wheel of our proposed method (i.e. equation~\ref{eq:combn}), we will find an analogy to scientific research that permuting grouping labels actually lists all the possible arrangements of observing the same random phenomenon and collecting two groups of data to reveal a differential pattern. However, in a single study, researchers observe the data once, and expect their data could convey the signal that the two groups are differential. To recover what might be missed, we propose progressive permutation and assume that the observed data are differential between the two groups. Then the method generates all the other disjointed possibilities in a systematic manner with fixed sample sizes so that the signal progressively accumulates from the full-permutation scenario to the no-permutation scenario. In other words, if the grouping factor is associated with the microbial difference between the two groups, the observed data defining the signals will be readily able to distinguish from the fully permuted data which characterizes the noise. Therefore, we achieve the identification of robust variables by judging that the significant p-values obtained from the observed data lie outside of the 95\% confidence intervals of the fully permuted data. 
}


%

\section*{Simulations}
In this section, we generate two types of simulations to show the performance of our method. First, we change the group mean, variance, correlation and number of significant variables to simulate data with different levels of signals. Second, we control the number of significant variables and simulate three data sets with different levels of heterogeneity. Then we compare the performance of our progressive permutation method on these data.

\begin{table}[h]
	\caption{Comparison on progressive permutation results produced by multiple simulated data sets with different simulation parameters, including correlation $\rho$, number of significant variables (nsv), group mean difference ($m_1-m_2$), and dispersion $\kappa$. AOI is short for area of interest. AUMC is short for area under the mixing curve. ``$\mathrm{slope}_0$'' denotes the slope of the first point in the U-Curve of number of significant features (the slope of the line connecting the first two points). ``$\mathrm{slope}_1$'' denotes the average value of the slope of the first 15 points ($K_f=15$) in the U-Curve of number of significant features (the slope of the line connecting the point with its next neighbor). ``fragility'' denotes the average value of the fragility index of the first 50 microbiome features. ``$\mathrm{select}_0$'' denotes the number of p-values that are less than 0.05 given by the testing results on the observed data. ``$\mathrm{select}_1$'' denotes the number of significant features identified by the proposed method.}
	\label{tab:simauc}
	\begin{footnotesize}
		\resizebox{\textwidth}{!}{\begin{tabular}{rrrrrrrrrrr}
				\cline{1-11}
				$\rho$ & nsv & $m_1-m_2$  & $\kappa$  & AOI & AUMC &  $\mathrm{slope}_0$ & $\mathrm{slope}_1$ & fragility & $\mathrm{select}_0$&$\mathrm{select}_1$ \\
				\cline{1-11}
				0.5 &   30 &    9 &	24 & 0.30 &	0.23 &	0 &	-0.54 &	6.64 & 30 & 30\\
				&     & 9&	1 &	0.30 &	0.20 &	0 &	-0.54 &	5.74 & 30 & 30\\
				&      &4 &	24 &	0.30 &	0.18 &	0 &	-0.52&	5.96 &30 &30\\
				&     &4 &	1 &	-0.05 &	-0.05 &	0.45 &	0 &	1.44 & 5 &3\\
				&       & 0 &	24	& 0.00 & 	-0.02 &	0 &	0.1	 & 0.44 & 0 & 0\\
				&       &0	& 1	 & 0.00	& -0.02	& 0	& 0.1 &	0.46 &0 & 0\\
				\cline{2-11} 
				&	90 &	9&	24&	1.00 &	0.67 &	0 &	-2 &	11.02 &100 &100
				\\
				& &9 &	1 &	1.00 &	0.58 &	0 &	-1.98 &	10.14 & 100 & 100\\
				&  &	4&	24 &	0.89 &	0.44&	-0.6 &	-1.74 &	8.84&89 &84
				\\
				& &4 &	1 &	-0.08 &	-0.08 &	1.5 &	-0.08 &	2.06&8 &3
				\\
				& &	0	& 24 &	0.00 &	-0.02 &	0 &	0.1 &	0.34&0 &0\\
				& &	0	& 1	& 0.00 & -0.02 & 0 & 0.1 &	0.5&0 &0\\
				\cline{1-11} 
				0.8	& 30 &	9&	24&	0.30&	0.23&	0&	-0.54&	6.58&30 &30\\
				&	&9&	1&	0.30&	0.20&	0&	-0.54&	5.8 &30 &30\\
				&  & 4 &24 &	0.30 &	0.18 &	0 &	-0.52 &	5.14&30 &30\\
				& & 4 &	1&	0.04&	0.04&	-0.45&	0.02&	1.2&4 &0 \\
				&&	0&	24&	0.00&	-0.02&	0&	0.1&	0.66& 0 &0\\
				& &	0&	1&	0.00&	-0.02&	0&	0.1 &	0.3&0 &0\\
				\cline{2-11} 
				&	90&	9	&24	&1.00&	0.67&	0&	-2&	11.12&100 & 100
				\\
				&&	9	& 1&	1.00&	0.58&	0&	-1.98&	10.1&100 &100\\
				&&	4	&24&	0.97&	0.45&	-2.1&	-1.9&	8.82& 97 & 87\\
				&&	4 &	1&	-0.07&	-0.08&	2.25&	-0.04&	2.06&7 & 0\\
				&&	0 &	24&	0.00&	-0.02&	0&	0.08&	0.54&0&0\\
				&&	0	&1 &	0.00&	-0.02&	0&	0.1&	0.2&0&0\\   
				\cline{1-11} 
		\end{tabular}}
	\end{footnotesize}
\end{table}

We follow the same simulation setup used by \cite{hawinkel2019broken}. We simulate the OTU counts as random samples drawn from a negative binomial distribution $ \mathcal{F}(m, \kappa)$, where $\kappa$ is called the dispersion parameter, as the variance is $m+\frac{m^2}{\kappa}$. To simulate the dependence between OTUs, we use the Gaussian copula~\cite{mcbook} to combine the correlation structure $\bm R$ with the negative binomial distributions. Here are the simulation steps. First, we draw Gaussian samples of $\bm Z\sim\mathcal{N}(0,\bm R)$. Second, we obtain the negative binomial samples $\bm X_j=\mathcal{F}^{-1}(\Phi(\bm Z_j)), j =1, \cdots, p$. $\Phi(\cdot)$ denotes the Gaussian cumulative distribution function. Third, we obtain the compositions by dividing each element $X_{ij}$ by a constant greater than the sum of each rows.

{
To gain a sense of how the shape of the U-Curve depicts the strength and robustness of signals, we construct multiple data sets, changing the simulation parameters and performing progressive permutation on each data set.  Let $x_{ij}^1 \sim \mathcal{F}(m_j^1, \kappa_j^1)$ denote the simulated data from Group 1. Let $x_{ij}^2 \sim \mathcal{F}(m_j^2, \kappa_j^2)$ denote the simulated data from Group 2. The two groups have the same sample size $n_1=n_2=30$ and the same correlation structure as $R_{ij}=\rho^{i-j}$. We simulate the grouping factor of interest $y$ as $[1,\cdots, 1, 2, \cdots, 2]$. Suppose both group consist of 100 variables. Let ``nsv'' denote the number of differential variables whose distribution means are $m_j^1$ or $m_j^2$, the means of all the other variables is set as 1. As shown in Table~\ref{tab:simauc}, we set the means of Group 1 as $\{10,10,10\}$ and the means of Group 2 as $\{1,6,10\}$, so the mean differences between the two groups are $\{9,4,0\}$.  For instance, a data set is generated with $m_1-m_2=9$ and nsv=$30$, meaning that 30\% of the 100 variables have strong differences ($m_j^1=10$ vs $m_j^2=1$, where $ j= 1, \cdots, 30$) between the two groups, while all the other 70 variables are not differential (mean difference is 0) between the two groups. We summarize the following observations based on the above simulations. AOI in general increases with the proportion of significant features in the simulated data. As the variance increases when $\kappa$ becomes smaller, the differential effect between the two groups shrinks with $\kappa$. So the AUMC and average fragility of the first 50 features become smaller. The differential effect increases with the two mean differences between the two groups. So the corresponding AUMC and average fragility of the first 50 features become smaller when mean differences are smaller. As shown in Figure~\ref{fig:examplecurve}, the shape of U-Curve becomes flatter when two groups are less differential. Therefore, the more a grouping factor differentiates the features, the bigger AOI, AUMC and fragility index will be obtained. In particular, when the mean difference between the two groups is close to 0, the AOI and AUMC are almost zero, indicating that the U-Curve of number of significant features is flat when there are no differential signals. Additionally, correlations between microbiome features do not affect the values of the AOI and AUMC. The significant features identified by the proposed method is a subset of features whose p-values are less than 0.05 in the observed data.
}

{However, the behavior of steepness of the U-Curve is not clear in the previous simulations. In the following simulations, we control the data to produce the same AOI, but with different slopes. In other words, the number of identified features are the same, but actually the robustness of these features are different. Rather than just consider the significance depicted by p-values, we can further consider robustness to evaluate the feature-outcome associations using the U-Curve and fragility index from progressive permutation. We will show that some unknown heterogeneity might be one reason affecting the robustness of the features that are identified as differential.} 
We generate three simulation data sets, which are denoted as SimData 1, SimData 2 and  SimData 3. They have the same sample size $n_1=n_2=30$ and same number of variables $p=100$. The 60 samples differ substantially between Group 1 (30 samples) and Group 2 (30 samples). We denote data of Group 1 as $D_1$ and data of Group 2 as $D_2$. For the 100 variables, we define the proportion of significant features to be 0.6, which implies that 60 variables are significant. To construct heterogeneity, we create the second source of difference by splitting Group 1 into two subgroups of samples, which are denoted as $D_{11}$ and $D_{12}$.  In the same way, we split Group 2 into two subgroups of samples, which are denoted as $D_{21}$ and $D_{22}$. The grouping factor of interest $y$ is $[1,\cdots, 1, 2, \cdots, 2]$. 

\begin{figure*}[!ht]
	\centering
	\includegraphics[width=0.95\linewidth]{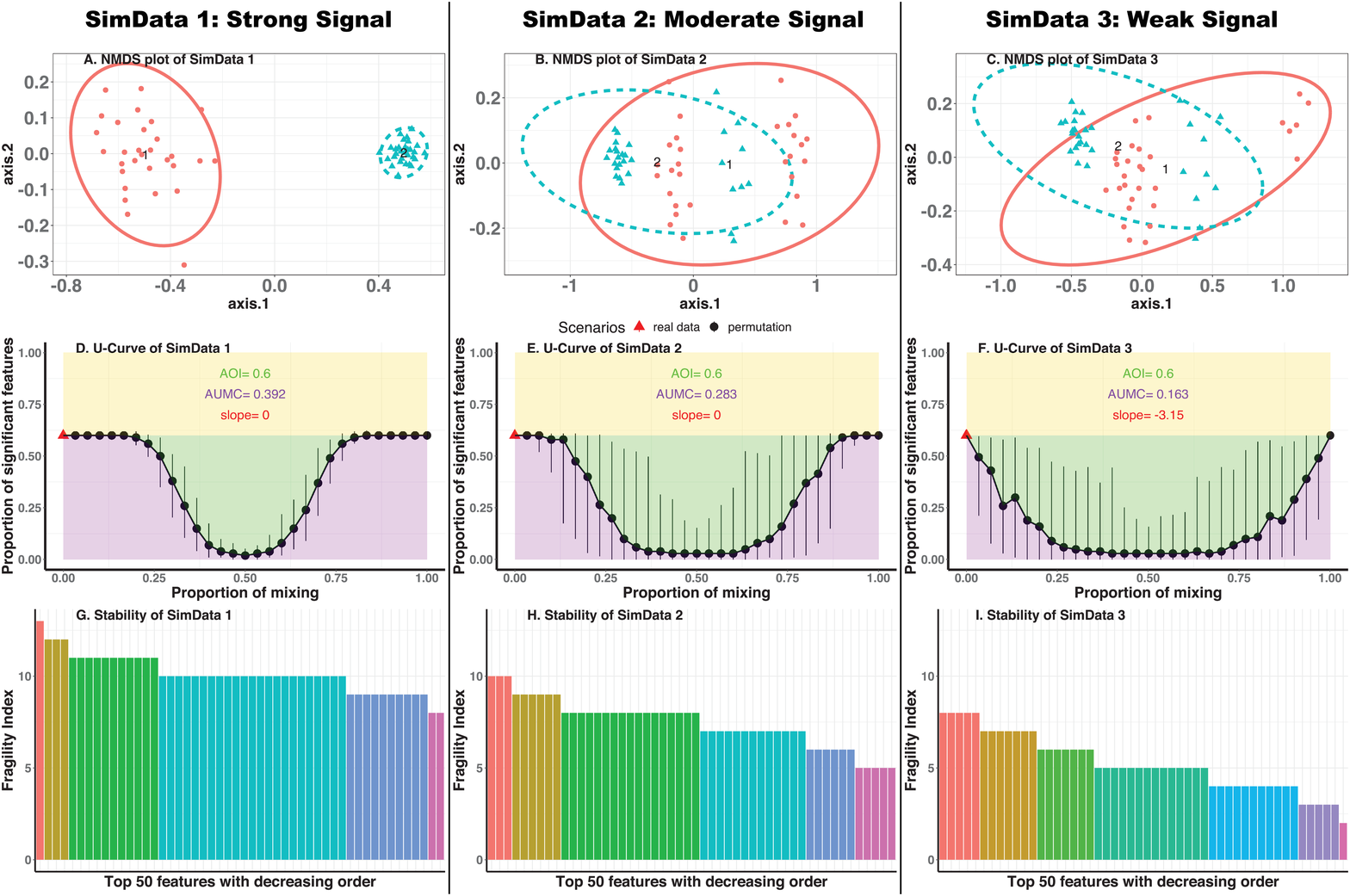}
	\caption{Result comparisons of three simulated data sets with different levels of heterogeneity. The first row (A,B C) shows the NMDS plot using the Bray-Curtis distance. The second row (D, E, F) shows the U-Curve of proportion of significant features. AOI is short for area of interest, which denotes the proportion of significant features out of all the features (area of green plus purple). AUMC is short for area under the mixing curve, which denotes the area under the U-Curve (area of purple). Slope denotes the slope of the red triangle. The red triangle denotes the real data. The third row (G, H, I) shows the fragility index. To save space, the legend listing the names of the 50 features are omitted.
	}
	\label{fig:simulation}
\end{figure*}

We describe the data generation as follows.  We use $(m)_c$ to denote a sequence containing $c$ number of $m$. $\mathrm{RN}(\mu_0, \sigma_0)$ describes the random number drawn from normal distribution with mean $\mu_0$ and variance $\sigma_0$. We define the correlation structure as $R_{ij}=\rho^{i-j}$.  $\rho$ is set up as 0.5.  Zero-Inflation is one of the main characteristics of microbiome data. Note that $\mu$ controls the magnitude of each variable and number of zeros in each sample. The distribution of zeros across samples and variables of SimData 1, SimData 2 and SimData 3  is comparable to the distribution of zeros in real Data, please see the histograms in Section S2 of Supplementary material. 

SimData 1: $D_{11}$ contains 8 samples. The mean is $[(6)_{30}, (4)_{30}, (1)_{40}]$.  The dispersion parameter $\kappa$ is 2.  $D_{12}$ contains 22 samples. The mean is $[(4)_{30}, (6)_{30}, (1)_{40}]$. The dispersion parameter $\kappa$ is 36. $D_2$ contains 30 samples. The mean is $[(15)_{30}, (0.5)_{30}, (1)_{40}]$. The dispersion parameter $\kappa$ is 36. 

SimData 2: $D_{11}$ contains 16 samples. The mean is $[(8)_{30}, (2)_{30}, (1)_{40}]$.  The dispersion parameter $\kappa$ is 25.  $D_{12}$ contains 14 samples. The mean is $[(2)_{30}, (8)_{30}, (1)_{40}]$. The dispersion parameter $\kappa$ is 24. $D_{21}$ contains 20 samples. The mean is $[(15)_{30}, (0.5)_{30}, (1)_{40}]$. The dispersion parameter $\kappa$ is 26.  $D_{22}$ contains 10 samples. The mean is $[(m_1)_{60},(m_2)_{40}]$, where $m_1= \mathrm{RN}(5, 1.2)$ and $m_2=\mathrm{RN}(1,0.1)$. The dispersion parameter $\kappa$ is 24. 

SimData 3: $D_{11}$ contains 24 samples. The mean is $[(8)_{30}, (2)_{30}, (1)_{40}]$.  The dispersion parameter $\kappa$ is 14.  $D_{12}$ contains 6 samples. The mean is $[(1)_{30}, (10)_{30}, (1)_{40}]$. The dispersion parameter $\kappa$ is 14. $D_{21}$ contains 20 samples. The mean is $[(15)_{30}, (0.5)_{30}, (1)_{40}]$. The dispersion parameter $\kappa$ is 14.  $D_{22}$ contains 10 samples. The mean is $[(m_1)_{60},(m_2)_{40}]$, where $m_1=\mathrm{RN}(5, 1.6)$ and $m_1=\mathrm{RN}(1, 0.3)$. The dispersion parameter $\kappa$ is 12.

Based on the above setup, we expect to see there are more and more levels of heterogeneity by constructing subgroups from SimData 1 to SimData 2 to SimData 3. As a result, the associations between the microbiome features and the grouping factor of interest is weaker and weaker because the proportion of differential samples between Group 1 and Group 2 is lower and lower. Traditionally, non-metric multidimensional scaling (NMDS) is used to collapse information from multiple dimensional features into just a few, so that clustering effect will be visualized and interpreted when we link them with a grouping factor of interest~\cite{cox2000multidimensional}. However, in the dimension reduction plots, the expected clustering effect can not be witnessed, because this main differential effect is mixed with heterogeneity. As shown in Figure~\ref{fig:simulation}, only the NMDS plot of SimData 1 shows us the clear cluster separations between Group 1 and Group 2. But both the NMDS plot of SimData 2 and the NMDS plot of SimData 3 show overlaps of Group 1 and Group 2 similarly. Therefore, NMDS plots could not distinguish the strength of the overall association between microbiome compositions and the grouping factor of interest. Besides, we can not visualize differences in heterogeneity between SimData 2 and SimData 3.

When testing the relationship between an explanatory variable and an outcome, the variable's effect might be modified by other variables and distorted by potential systematic bias, confounding or effect modification. The U-Curve and fragility index plots provides us with a measure of all these disturbances mixed with the main signals in the collected data. The U-Curve provides a dynamic depiction of how our method progressively singles out signals from randomized trials. In each plot, the number of significant features decreases from observed data to full permutation scenario. The shape becomes steeper when the associations are less stable (with more disturbances). We use AUMC (area under the mixing curve) to quantify the shape of the U-Curve. AUMCs in Figure.~\ref{fig:simulation}D, E and F are 0.392, 0.283 and 0.163, which ranks the decreasing order of robustness of the association between microbiome compositions and the grouping factor. The average fragility index of the top 50 features are 10.12 for SimData 1, 7.44 for SimData 2, and 5.24 for SimData 3. Since the full permutation scenario $K_f=15$, the average scaled fragility indices are 0.675 for SimData 1, 0.496 for SimData 2 and 0.349 for SimData 3. 

Please note that, 
when generating the U-Curve plots (D, E, F in Figure~\ref{fig:simulation}), the black dots describe the median value. The black bars describe the $2.5\%$ and $97.5\%$ quantile intervals. We follow the same setup in all the subsequent figures.

\section*{Application}

In this section, we apply the proposed method into two microbiome studies. The first study includes five groups. We regroup them to construct two data sets with different levels of heterogeneity. In the second study, we link microbiome compositions with two different outcomes. 

\begin{figure*}[!ht]
	\centering
	\includegraphics[width=0.95\linewidth]{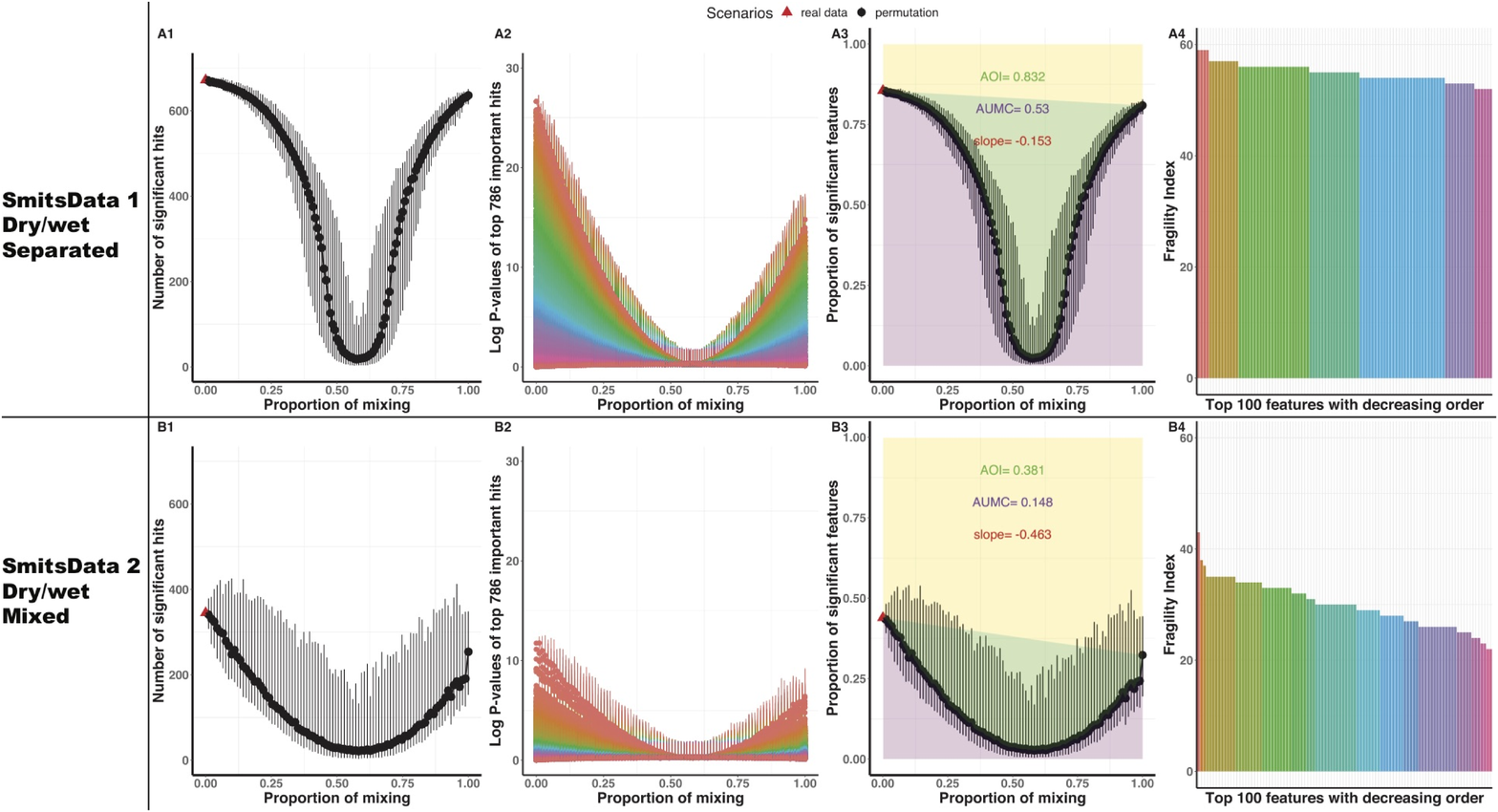}
	\caption{Result comparisons of regrouped data SmitsData 1 (A1-A4) and SmitsData 2 (B1-B4) with different levels of heterogeneity. A1 and B1 plot the U-Curve of number of significant hits. In A2 and B2, we rank the significance of the 786 features, and then plot their $-\log_{10} p$-values with the same order across permutation scenarios.  A3 and B3 plot the U-Curve of proportion of significant hits. A4 and B4 plots the fragility index of the top 100 features with a decreasing order. To save space, the legend listing the names of the 100 features are omitted.
	}
	\label{fig:hadza_data}
\end{figure*}

\begin{figure*}[!ht]
	\centering
	\includegraphics[width=0.95\linewidth]{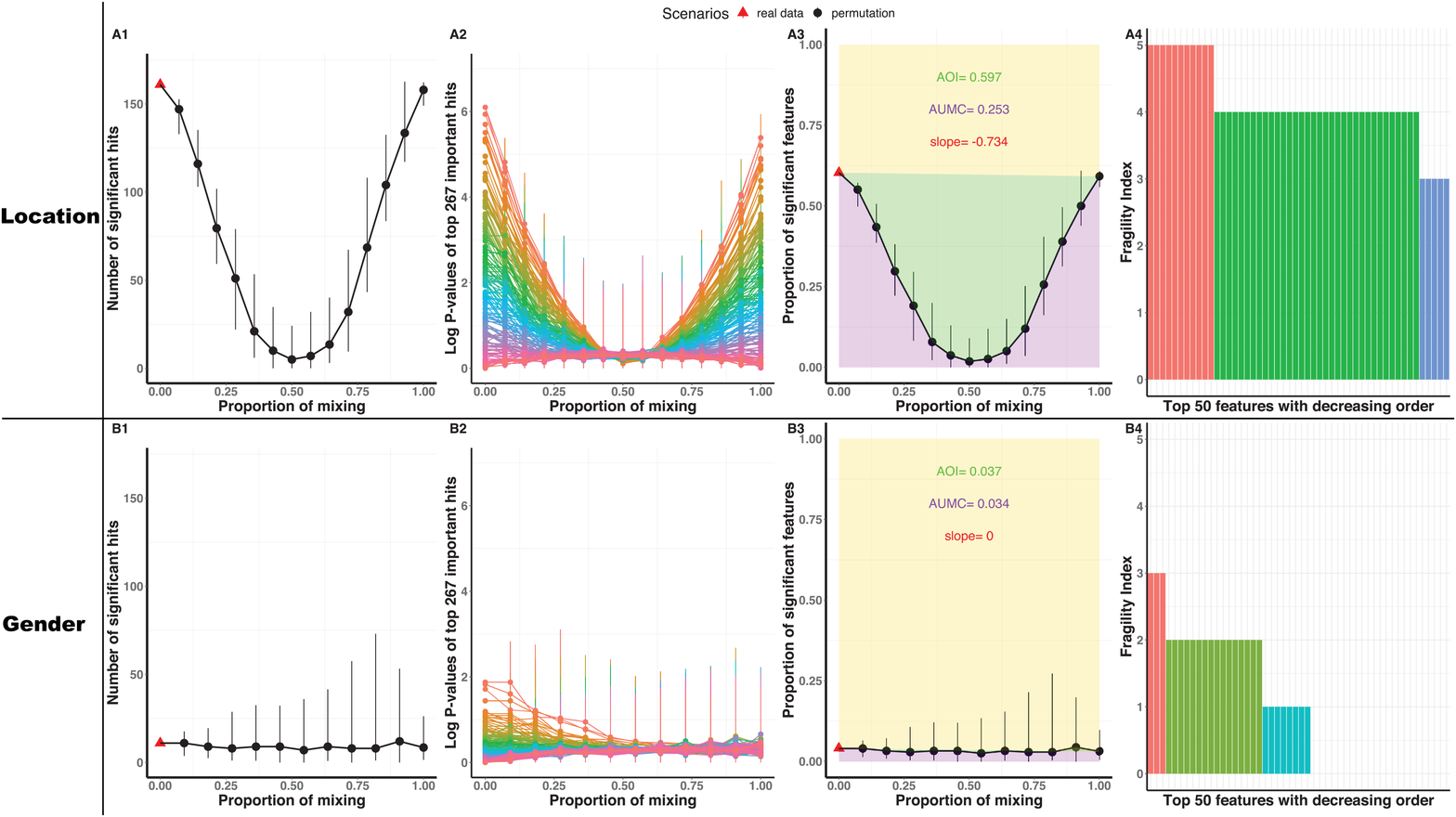}
	\caption{Result comparisons when linking microbiome compositions with location (A1-A4) and gender (B1-B4). A1 and B1 plot the U-Curve of number of significant hits. In A2 and B2, we rank the significance of the 267 features, and then plot their $-\log_{10} p$-values with the same order across permutation scenarios. A3 and B3 plot the U-Curve of proportion of significant hits. A4 and B4 plots the fragility index of the top 50 features with a decreasing order. To save space, the legend listing the names of the 50 features are omitted.
	}
	\label{fig:italian_child}
\end{figure*}

The first study examined the gut microbiota  of 350 stool samples collected longitudinally for more than a year from the Hadza hunter gatherers of Tanzania. The samples were collected subsequently with 5 seasonal groups: 2013-LD (Late dry), 2014-EW (Early wet), 2014-LW (Late wet), 2014-ED (Early dry) and 2014 LD (Late Dry).  Smits SA, et al.~\cite{smits2017seasonal} found that Hadza gut microbial community compositions are cyclic and differential by season. They observed that samples from the dry seasons were distinguishable from the wet seasons and were indistinguishable from other dry seasons in sequential years.  
We combine 2014-ED ($n=33$) and 2014-LD ($n=133$) as the ``Dry'' group, and combine 2014-EW ($n=62$) and 2014-LW ($n=58$) as the ``Wet'' group. We call this regrouped data as SmitsData 1. In the same way, we combine 2013-LD ($n=64$) and 2014-EW ($n=62$) as the ``LDEW'' group, and combine 2014-LW ($n=58$) and 2014-ED ($n=33$) as the ``LWED'' group. We call this regrouped data as SmitsData 2. We expect that SmitsData 1 is more differential between Dry and Wet group than SmitsData 2 between LDEW and LWED group.

{
In total, we have 786 taxonomic features. We perform the progressive permutation tests on SmitsData 1 (Dry $n_1=166$ vs. Wet $n_2=120$) and SmitsData 2 (LDEW $n_1=126$ vs. LWED $n_2=91$). The results of SmitsData 1 (A1-A4) and SmitsData 2 (B1-B4) are shown in Figure~\ref{fig:hadza_data}.  In the observed data (no permutation), differential tests provide more significant hits (p-value less than 0.05) from SmitsData 1 (672 in A1) than SmitsData 2 (345 in B1). There are more $-\log_{10} p$-values greater than $-\log_{10} 0.05$ (A2 vs. B2). The U-Curve of SmitsData 1 (AUMC is 0.53) is steeper than SmitsData 2 (AUMC is 0.148). Based on the plot of fragility index, the overall robustness of the top 100 features from SmitsData 1 (average fragility index is 54.93 in A4) is more than SmitsData 2 (average fragility index is 29.93 in B4). The full permutation scenario for SmitsData 1 is $K_f=70$. So the average scaled fragility index for SmitsData 1 is 0.785. The full permutation scenario for SmitsData 2 is $K_f=53$. So the average scaled fragility index for SmitsData 2 is 0.565. In addition, the initial slopes of the first points for SmitsData 1 and SmitsData 2 are -0.153 and -0.463 respectively, which also indicate the significance in SmitsData 1 is more robust. All these results demonstrate that the progressive permutation results can convey and quantify the overall association which is disturbed by heterogeneity. When it comes to feature identification, the proposed method obtains 656 features for SmitsData 1 and 271 features for SmitsData2.}

The second study investigated the impact of diet by comparing the gut microbiota of 14 children aged 1-6 years in a village of rural Africa with the gut microbiota of 15 European children of the same age. De Filippo C, et al.~\cite{de2010impact} found significant differences in gut microbiota between the two groups, as children at these two locations have different dietary habits. 11 of them are female. 18 of them are male. There is almost no difference in microbiome compositions by gender. In total, we have 267 taxonomic features. We perform the progressive permutation tests to associate microbiome compositions with location and gender respectively. {The results of location (A1-A4) and gender (B1-B4) are shown in Figure~\ref{fig:italian_child}. In the observed data, differential tests provide more significant hits for Location (161 in A1) than for Gender (11 for A2). The results illustrate that microbiome compositions are strongly associated with location instead of gender, because AUMC of location (0.253 in A3) is greater than AUMC of gender (0.035 in B3). The U-Curves of gender (B1 and B3) are almost flat, which imply that the overall association between microbiome compositions and gender is weak. Based on the plot of fragility index, the overall robustness of the top 50 features for Location (average fragility index is 4.12 in A4) is more than Gender (average fragility index is 0.98 in B4). The full permutation scenario for Location is $K_f=7$, and the average scaled fragility index for SmitsData 1 is 0.589. The full permutation scenario for gender is $K_f=7$, and the average scaled fragility index for gender is 0.14. In addition, the average slopes of the first 7 points for location and gender are -1.17 and -0.03 respectively, which also indicate there is no significance for gender across all the scenarios. All these results demonstrate that the progressive permutation method can measure and rank the overall association between microbiome and multiple outcomes of interest. For the outcomes with high association, we will continue to identify the microbiome features that are linked to them. }

\begin{figure*}[!ht]
	\centering
	\includegraphics[width=0.95\linewidth]{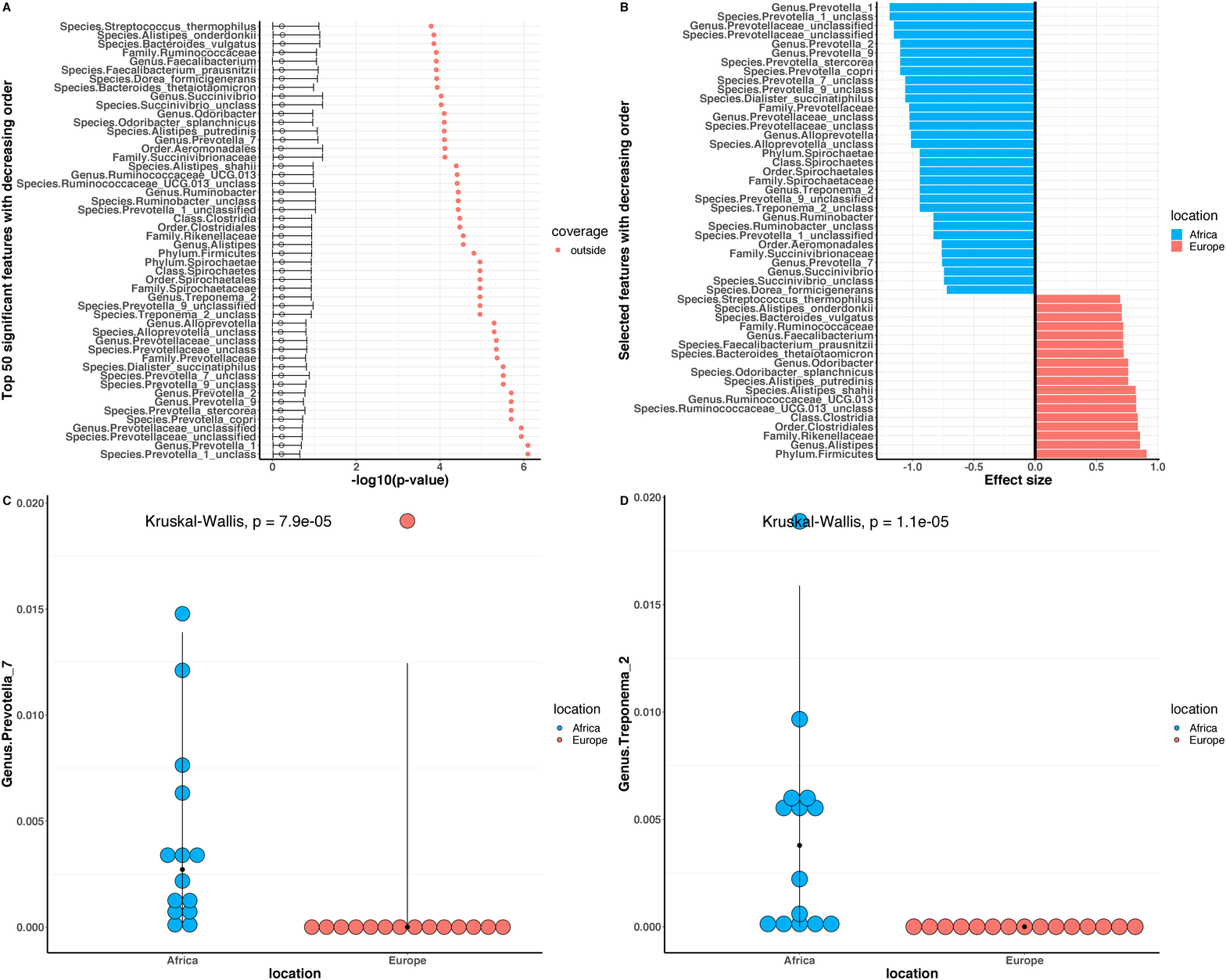}
	\caption{List of discoveries, effect sizes and individual abundances. A denotes the coverage plot of the top 50 features with decreasing order. The color dots denote the $-\log_{10} p$-value of top 50 features in the original data (permutation proportion is 0). The horizontal bars describe the 95\% quantile confidence intervals of the $-\log_{10} p$-value in the full permutation scenario. B denotes the effect sizes of identified features. C and D denote the dot plot of abundance of Prevotella and Treponema with median-quantile vertical lines.
	}.
	\label{fig:Children_identify}
\end{figure*}

We include the identification of individual features in our software by observing whether the $-\log_{10} p$-values of targeted features lie within the 95\% confidence interval of median $-\log_{10} p$-values of the full mixing scenario. The proposed method has identified 155 features for location and 0 features for gender. As shown in the upper left panel in Figure~\ref{fig:Children_identify}, all the top 50 features are significant. The effect sizes of these 50 significant features are plotted in the upper right panel.  Our findings are consistent with published results~\cite{de2010impact}. Firmicutes are more abundant in European children than in  African children.  Prevotella and Treponema (Spirochaetaceae) are more abundant in African children than in European children (as shown in the lower panels of Figure~\ref{fig:Children_identify}). 

In summary, our method first explores the overall association (that might be complicated by heterogeneity) between microbiome compositions and outcome variable. If the association is reasonable, it will identify the significance of individual hits, list their effect sizes and plot individual abundances.

\section*{Analytical property}


Various summary statistics, like mean, variances, median and rank sums, have been used to analyze differences between two groups. Each statistic goes along with an assumption of a sample distribution, including normal, negative binomial and so on.  Among these, the mean test under a normal assumption is one of the most widely-used statistical techniques for group comparisons. Other types of tests extend the standard to broader situations that require specific assumptions or less restrictions. Therefore, it is worthwhile to pursue the theoretical aspects of the progressive permutation method in a basic setup that performs Z-tests. The theoretical results from parametric tests can provide insights to the progressive permutation coupling non-parametric tests, as we expect to observe similar patterns between them. To simplify the problem, we assume observing two groups of variables from Gaussian family. Both groups have the same number of variables $p$. The population distribution of Group 1 is $\mathcal{N}(\mu_j^1,\sigma^2)$, and the population distribution of Group 2 is $\mathcal{N}(\mu_j^2,\sigma^2)$. We aim to test the hypothesis $H_{0j}: \mu_j^1=\mu_j^2,\ vs.\ H_{1j}: \mu_j^1\neq \mu_j^2$. We use $x_{ij}^1$ to denote the $i$th observation of the $j$th variable in Group $1$ and $x_{ij}^2$ to denote the $i$th observation of the $j$th variable in Group $2$. We simulate the data from Gaussian distributions with $x_{ij}^1\sim\mathcal{N}(m_j^1,\sigma^2)$ and $x_{ij}^2\sim\mathcal{N}(m_j^2,\sigma^2)$. The observations of every variable in each group are independent and identically distributed. 
We denote the grouping labels in Group 1 as $I^1=\{1, \cdots, n_1\}$. We denote the grouping labels in Group 2 as $I^2=\{1, \cdots, n_2\}$. 

\begin{figure}[!ht]
	\centering
	\includegraphics[width=0.9\linewidth]{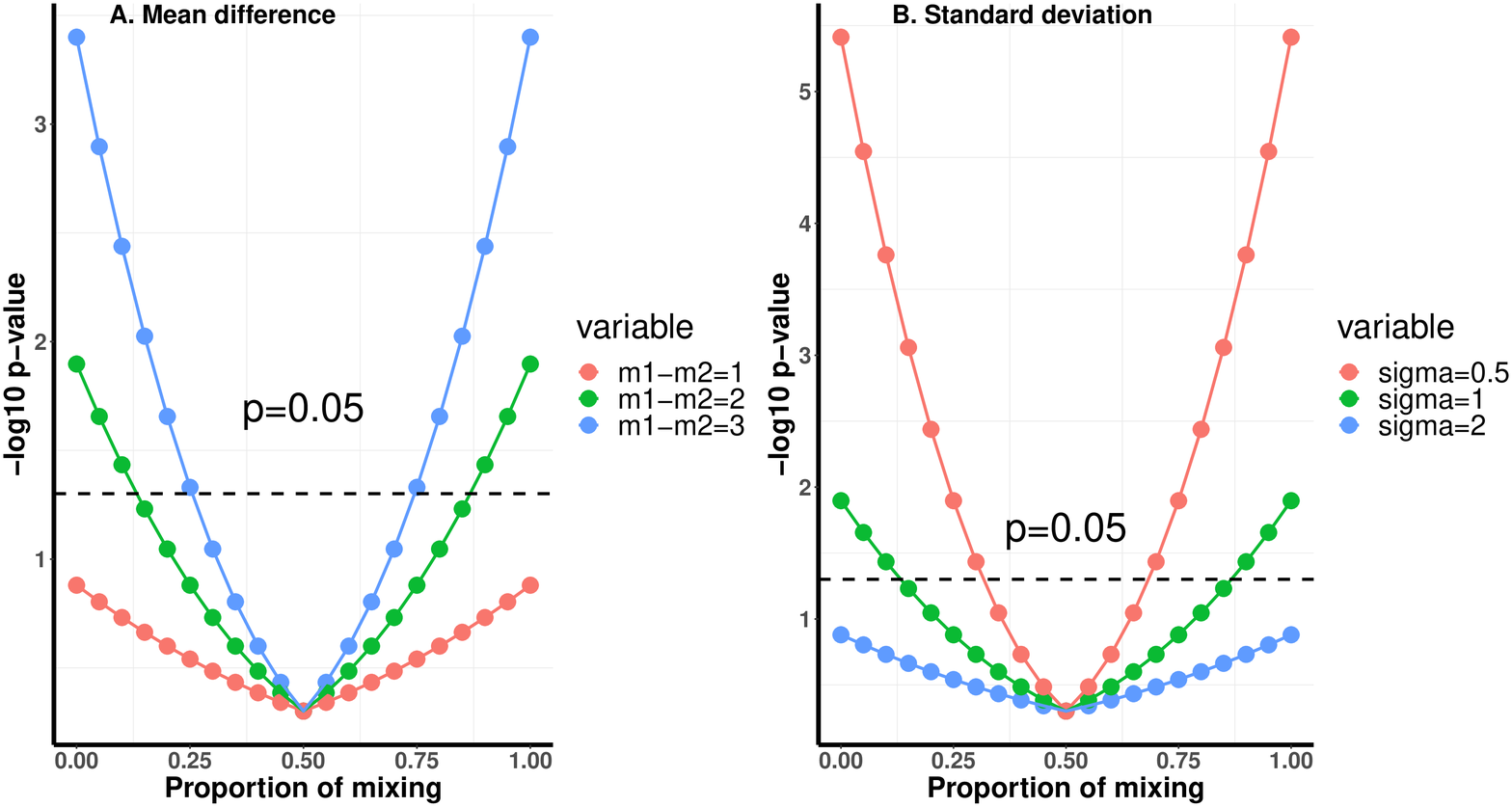}
	\caption{U-Curve plots of p-values calculated from formula (\ref{eq:pvalue}). Both of the sample sizes $n_1$ and $n_2$ are 20. x-axis is $\frac{k}{K}$.  In panel A, standard deviation $\sigma$ is fixed at 2 . In panel B, mean difference $m_j^1-m_j^2$ is fixed at 1. }
	\label{fig:theory_mu_sigma}
\end{figure}

To test the difference of the $j$th variable between the two groups of the original data, we calculate the mean difference between the two groups, 
\begin{equation}\label{eq:original_test}
\bm \bar{x}_j^1-\bm \bar{x}_j^2=\frac{1}{n_1}\sum_{i=1}^{n_1} x_{ij}^1-\frac{1}{n_2}\sum_{i=1}^{n_2} x_{ij}^2\sim \mathcal{N}(m_j^1-m_j^2,\frac{n_1+n_2}{n_1n_2}\sigma^2).
\end{equation}
Suppose we perform the progressive permutation method and randomly draw $k$ samples from group 1 and $k$ samples from group 2, and then exchange their grouping labels. We denote the selected labels in Group 1 as $I_k^1=\{i^1_1, \cdots,  i^1_k\}$. We denote the selected labels in Group 2 as $I_k^2=\{i^2_1, \cdots, i^2_k\}$. Then the mean difference of the $j$th variable between the two groups of permutation scenario $k$ becomes 
\begin{equation}\label{eq:permute_test}
\begin{split}
\bm \bar{x'}_j^1-\bm \bar{x'}_j^2=&\frac{1}{n_1}\sum_{i\in I^1\setminus I^1_k} x_{ij}^1 + \frac{1}{n_1}\sum_{i\in I^2_k}x_{ij}^2
- \frac{1}{n_2}\sum_{i \in I^2\setminus I^2_k} x_{ij}^2 -  \frac{1}{n_2}\sum_{i\in I^1_k}x_{ij}^1\\
\sim& \mathcal{N}\left((1-\frac{n_1+n_2}{n_1n_2}k)(m_j^1-m_j^2),\frac{n_1+n_2}{n_1n_2}\sigma^2\right).
\end{split}
\end{equation}
We assume $m_j^1>m_j^2$. The differences of sample means after permutation (\ref{eq:permute_test}) are smaller than those before permutation (\ref{eq:original_test}). 
Suppose $\delta_j=\frac{m_j^1-m_j^2}{\sigma}$. The p-value of the $j$th variable is
\begin{equation}\label{eq:pvalue}
\begin{split}
p_j(k)&= P\left(|z|>\frac{(\bm \bar{x'}_j^1-\bm \bar{x'}_j^2)-(\mu_j^1-\mu_j^2)}{\sqrt{\frac{n_1+n_2}{n_1n_2}\sigma^2}}\bigg\vert H_{0j}\right)\\
&=2P\left(z+\frac{\bm \bar{x}_j^1-\bm \bar{x}_j^2}{\sqrt{\frac{n_1+n_2}{n_1n_2}\sigma^2}}<0\right)\\
&= 2\Phi\left(-\sqrt{\frac{n_1n_2}{2(n_1+n_2)}}(1-\frac{n_1+n_2}{n_1n_2}k)\delta_j\right),
\end{split}
\end{equation}
where $k\leq \lceil \frac{n_1n_2-1}{n_1+n_2+2} \rceil$.  $\Phi(\cdot)$ denotes the cumulative function of standard normal distribution. Therefore, with the increase of exchanged labels $k$, $-\log_{10} p$-value is smaller. 
As we perform two sided Z-tests in each scenario, the permutation results (p-values) are symmetric with respect to the fully mixing scenario $K_f=\lceil \frac{n_1n_2-1}{n_1+n_2+2} \rceil$. Then we can obtain the p-value of the $j$th variable when $k=K_f, \cdots, K$ as $p_j(k)=2\Phi\left(\sqrt{\frac{n_1n_2}{2(n_1+n_2)}}(1-\frac{n_1+n_2}{n_1n_2}k)\delta_j\right)$. $-\log_{10} p_j(k)$ decreases with $k$ when $0\leq k\leq K_f$ and increases with $k$ when $K_f\leq k\leq K$. {If we generate the simulated data under the null hypothesis meaning that $\delta_j=0$, all the $p$-values will be 1 across all permutation scenarios so that the curve of $-\log_{10} p$-values and number of significant features will become a flat horizontal line at 0.}  We define $\frac{k}{K}$ as the proportion of mixing. We let $n_1=n_2=20$. {As we generate the simulated data under the alternative hypothesis meaning that $\delta_j>0$, then we can observe in Figure~\ref{fig:theory_mu_sigma}, $-\log_{10} p_j(k)$ is a U-Curve of $\frac{k}{K}$.} If the differences of sample means are bigger, the U-Curve is steeper. If the standard deviation of the samples is bigger, the U-Curve is flatter. Therefore, the shape of the U-Curve measures how differential the quantifies of interest are between the two groups. 

\section*{Discussion}
{The proposed method considers the signal identification problem as progressively singling out signals from permuted randomized versions of an original data set. It progressively permutes the grouping factor of the microbiome data and performs multiple differential abundance tests in each scenario. We then summarize the resulting p-values by the number of significant hits and calculate their fragility index to convey the overall association with the grouping factor and robustness of the discovered microbes. Based on these global characteristics, we can rank grouping factors by the strength of association with microbiome and then decide the necessity of identifying the significance of microbiome features that are linked with the targeted grouping factor. We achieve the identification of microbiome features by comparing the p-values of the observed data with the confidence region of p-values of the fully permuted data. We have developed these methods into a user-friendly and efficient R-shiny tools with visualizations.}

Our proposed method will present a U-Curve when the overall association between microbiome and grouping factor is not zero. Furthermore, it can quantify different levels of heterogeneity which is a common phenomenon in medical research. The associations between microbiome compositions and a grouping factor may be weak due to a variety of reasons, such as systematic bias, confounding and effect modification. Therefore, we generate three simulated data sets with different levels of heterogeneity. The AUMC and fragility index provide a summary of the progressive permutation results to quantify and differentiate heterogeneity. In addition, the application into Smits data and DeFilippo data continue to prove that our proposed method can convey the overall association between microbiome compositions and outcome of interest, rank the robustness of the discovered microbes and identify robust individuals. 

In this paper, we link microbiome composition with a binary outcome. It creates a new concept and framework to understand significance and robustness of identified microbiome features. Following the same logic, we can extend the binary outcome to a continuous outcome. When constructing the progressive permutation scenarios, we permute a proportion (select $k$ samples and calculate $\frac{k}{n}$) of the continuous outcome. In each scenario, we perform the Kendall's Tau and Spearman's Rank Correlation tests to associate microbiome compositions with the permuted continuous outcome. We then adopt similar procedures of the binary outcome to summarize the permutation results. We have applied the progressive permutation with a continuous outcome to a sample data set. Please see the results in Section S3 of the Supplementary material. 

{Our paper is mainly designed for the exploration stage of microbiome data analysis, where we provide visualizations to illustrate the robustness of potential hits. Researchers could take the results generated from our method as a guide to help with the final decision on feature identification. Notably our method does not calculate adjusted p-values. In the future, we will consider to use progressive permutation results to estimate the null distribution to correct false positive rates, so we can adjust the existing p-values according to the null distribution and identify robust hits based on the adjusted p-values. }


\begin{backmatter}
	
\section*{Abbreviations}
AOI: area of interest; AUMC: area under the mixing curve.

\section*{Acknowledgements}

The authors thank Professor J. Jack Lee for providing us great suggestions on simulations,
Norris Clift for helping us launch the Shiny App on the server https://biostatistics.mdanderson.org/shinyapps/ProgPerm.

\section*{Authors' contributions}

RJ conceived the study. RJ, CP and KD provided guidance for this project. LZ and RJ proposed the method. LZ implemented the simulation and methods. LZ analyzed the result with the help of YS, KD, CP and RJ. LZ wrote the manuscript with the help of CP and RJ.

\section*{Funding}

KAD is partially supported by MD Anderson Moon Shot Programs, Prostate Cancer SPORE P50CA140388, NIH/NCI CCSG grant P30CA016672, CCTS 5UL1TR000371, and CPRIT RP160693 grants. CBP is partially supported by NIH/NCI CCSG grant P30CA016672 and MD Anderson Moon Shot Programs. RRJ is partially supported by NIH R01 HL124112 and CPRIT RR160089 grants.

\section*{Availability of data and materials}
RShiny App is accessible at \url{https://biostatistics.mdanderson.org/shinyapps/ProgPerm}. R codes and example data are available at \url{https://github.com/LyonsZhang/ProgPerm}.

\section*{Ethics approval and consent to participate}
Not applicable.

\section*{Consent for publication}
Not applicable.

\section*{Competing interests}
The authors declare that they have no competing interests.


\bibliographystyle{vancouver} 
\bibliography{liangref}      

\section*{Additional file 1}
Supplementary document. Section S1: Mathematical notations.  Section S2: Distribution of zeros. Section S3: Results of continuous outcome. 
%

\end{backmatter}

\end{document}